\documentclass[12pt,a4paper]{article}
\usepackage[utf8]{inputenc}
\usepackage[T1]{fontenc}
\usepackage{geometry}
\usepackage{array}
\usepackage{multicol}
\usepackage{comment}
\usepackage{amsopn,amsmath,amssymb,amsthm,amstext,amsfonts,algorithmic,psfrag}
\usepackage{amssymb}
\usepackage{enumitem}
\usepackage{mathtools}
\usepackage[english]{babel}
\usepackage{hyperref}
\hypersetup{colorlinks=true, citecolor=blue, linkcolor=blue}
\usepackage[round]{natbib}
\setcitestyle{aysep={,}} 
\usepackage{authblk}
\usepackage{graphicx}
\usepackage{mwe}    
\usepackage{subfigure} 
\usepackage{float}
\usepackage{caption}
\usepackage{algorithm}
\usepackage{algorithmic,latexsym}
\usepackage{placeins,enumitem}
\renewcommand{\thetable}{\arabic{table}}

\usepackage{eso-pic}
\newcommand\AtPageUpperMyright[1]{\AtPageUpperLeft{%
 \put(\LenToUnit{0.5\paperwidth},\LenToUnit{-1cm}){%
     \parbox{0.5\textwidth}{\raggedleft\fontsize{9}{11}\selectfont #1}}%
 }}%
\newcommand{\conf}[1]{%
\AddToShipoutPictureBG*{%
\AtPageUpperMyright{#1}
}
}

\title{\LARGE \bf
Optimal Lockdown Management using  Short Term COVID-19 Prediction Model
}

\author[1]{Shuvrangshu Jana}

\author[2]{Debasish Ghose}

\affil[1]{Post-doctoral Fellow, Department of Aerospace Engineering, Indian Institute of Science, Bangalore,
		{\tt\small shuvrangshuj@iisc.ac.in}}
		
\affil[2]{Professor, Department of Aerospace Engineering, Indian Institute of Science, Bangalore, 
		{\tt\small dghose@iisc.ac.in}}		
		
\date{}                     
\renewcommand{\baselinestretch}{1.5}
\begin{document}
\maketitle
\conf{$5^{th}$ World Congress on  Disaster Management\\ IIT Delhi, New Delhi, India, 24-27 November 2021}


\begin{abstract}
This paper proposes optimal lockdown management policies based on  short-term prediction of active COVID-19  confirmed cases to ensure the availability of critical medical resources.  The optimal time to start the lockdown from the current time is obtained after maximizing a cost function considering economic value subject to constraints of availability of medical resources, and maximum allowable value of daily growth rate and Test Positive Ratio. The estimated value of required medical resources is calculated as a function of total active cases. The predicted value of active cases is calculated using an adaptive short-term prediction model.  The proposed approach can be easily implementable by a local authority. An optimal lockdown case study for Delhi during the second wave in the month of April 2021 is presented using the proposed formulation. 
\end{abstract}


\section{Introduction}

  Lockdown is one of the non-pharmaceutical intervention tools to manage the spread and impact of  COVID-19. Lockdown provides time for building  new medical infrastructure and slows down the growth of active cases; however,  it affects the economic condition of that region. Optimal management of lockdown is necessary to balance the medical cost in terms of human life and the financial cost associated with shutting down economic activities.

 Management of lockdown consists of three phases: a) Decision of lockdown, b)  Decision of lockdown duration, c)  Lifting of lockdown.    Several works have been reported in the literature related to optimal management of lockdown \citep{gonzalez2020optimal,acemoglu2020multi,federico2021taming}.  Most of these results try to determine the optimal duration of lockdown considering the physical disease model and economic model.  Initiating lockdown is an important decision for a government authority as it is associated with  economic \citep{singh2020income,mottaleb2020covid}, medical, social \citep{dubey2020psychosocial,usher2021covid,kumar2021migrant}, and political cost \citep{ren2020pandemic}. The quantitive modeling of these costs is difficult, and sometimes there is a trade-off between different costs. So, the government authority needs to balance the different costs to find an optimal solution for the decision of starting lockdown \citep{alvarez2020simple, miclo2020optimal}.   Mathematical modeling of different costs is not feasible; therefore, decision-making authorities generally lack a quantitative tool to justify the decision of lockdown. Also, it might be practically impossible to develop this cost model dynamically for developing countries or for a small region.

 Methodology for the decision of lockdown is reported in literature, such as \citep{hikmawati2021multi,li2020novel,ghosh2021mathematical,alvarez2020simple, bandyopadhyay2020learning}. Lockdown strategy using optimal control, and SIER model is reported in  \citep{rawson2020and}; however, it will be difficult to obtain the model parameters for each region.    \cite{li2020novel} proposes lockdown when the test capacity is 16 times the new cases based on the early-stage information from the many Italian regions such as  Marche and Lazio; however, this criterion does not consider the status of availability of medical resources and medical infrastructure.

 In \citep{alvarez2020simple}, the decision for lockdown is formulated as an optimal control problem using the SIR model.  \cite{hikmawati2021multi} proposes recommendation based on multi-criteria from multiple databases of different information such as population number, population density, ratio of elderly population, total cases, health facilities, health workers, susceptible population, etc. Local authorities should consider lockdown decisions for better efficiency; therefore, lockdown criteria formulation based on the physical model will have limited application as the model of most areas will not be available. 
 
 The important factor for the decision of lockdown is the estimation of required resources over a future horizon.  The requirement of critical resources has a good correlation with the current status of dynamics of the pandemic; therefore, the prediction of dynamics of the pandemic is crucial for lockdown decision making. Prediction of future cases are reported in several literature using physical modeling \citep{rajesh2020covid,huang2020global}, data-driven approach \citep{bertsimas2021predictions,chakraborty2020real,jana2020adaptive} or hybrid methods \citep{barmparis2020estimating, gupta2020estimating}. A good lockdown algorithm should avoid the requirement of obtaining the physical parameters of dynamics, as these parameters are highly uncertain.

 In this paper, the lockdown policy is formulated without using any physical model, and it ensures the availability of critical medical resources while maximizing economic activity.  The effect of lockdown on the total number of active cases is visible only after a certain duration of imposing the lockdown. This duration is approximately observed as two weeks.  This time lag to see the effect of lockdown is considered in the proposed formulation.  The requirement of medical resources is related to the total number of active cases.  We have predicted the requirement of medical resources as a function of the predicted value of active cases, and lockdown criteria are formulated to ensure that requirement of any resources should not exceed the availability at any time.  The active cases are predicted using an adaptive short-term prediction model based on the observations of total confirmed cases, recovery cases, and total death cases.   The prediction model can be used for varied population,  and area, and it only requires the time series data of total confirmed cases, recovered cases, and death cases.  The prediction model can be used to predict over the next 3-4 weeks with good accuracy.   The algorithm needs to be updated with daily new observations, and a lockdown policy can be formulated dynamically.
 
 We have considered a case study for the city of Delhi to derive optimal lockdown criteria during its second wave's rise on April 2021. It is found that Delhi had availability of less amount of oxygen than the requirement at a certain period. The case study is formulated with reasonable assumptions to obtain the optimal lockdown decision such that the shortage of oxygen could have been avoided.  In the particular scenario,  it is found that the implementation of lockdown a week before the actual implementation could keep the requirement of oxygen within the available limit.

 The rest of the paper is described as follows:   Section II describes the different criteria for the decision of lockdown.  The mathematical formulation of the lockdown policy is described in Section III. A case study for optimal policies for lockdown for a city is presented in Section  IV.

 \section{ Lockdown criteria} 
 Lockdown reduces the load on the medical infrastructure; however, it has a huge economic cost and social consequences. Therefore, lockdown needs to be managed optimally considering social, medical, and economic factors.
 Lockdown decision is reported in the literature based on various factors such as population number, population density, ratio of elderly population \citep{hikmawati2021multi}. 
  
   Criteria of the decision of lockdown will depend on the stage of the pandemic, dynamics of disease,  availability of critical medical resources, load on the medical infrastructures,  economic impact, and social factors.  Many of these factors can be modeled quantitatively. 

  In the early stages of a pandemic, lockdown is employed to stop the spread of disease among the population, and it can be decided based on the total number of cases. Availability of medical items over a large future horizon could influence the decision; however, isolating the few infected cases is the main criteria to lockdown in the initial pandemic stages.   In the early stages, lockdown can be a central decision for a shorter duration; however, the criteria for the lockdown need to be decided by the local authority based on local factors.
  
  At the middle stage of the pandemic, apart from controlling the growth rate of infection, ensuring the availability of medical resources should be the main criteria for the decision of lockdown.  So, in general, The critical factor behind the decision of lockdown by the government authority is the availability of medical items and the load on medical infrastructure.  
 
 The current dynamics of the disease are measured by the growth rate of total cases, and to some extent, by the test positive ratio.  Although the number of testing has a  considerable impact on the test positive ratio, the  test positive ratio has a positive correlation with the spread of disease.  The basic reproduction number is also one of the parameters which reflects the rate of the pandemic. 
 
 Sometimes the storage and distribution of resources could be a bottleneck to ensure the availability of resources for every patient.  Some items like oxygen require special vehicles to supply from resource points to hospitals, and this could be a factor if sufficient special vehicles are not available. Also, some areas might not have sufficient storage capacity. The coupled factor of storage and distribution could severely hinder the availability of resources.  During the second wave of COVID-19  in India, it is observed that the storage and distribution of oxygen are crucial factors for the successful management of COVID-19. 
 
 The following considerations are used for the decision of lockdown. 
 
 \begin{enumerate}
     \item  Total number of COVID beds (ICU, oxygen)
     \item  Availability of critical medical items (medicine, oxygen, ventilator) 
     \item  Test positive ratio
     \item  Growth rate of total cases
     \item  Distribution capability
 \end{enumerate}

 \section{Mathematical formulation}
 In this section an optimization framework is proposed  to decide the optimal duration  after  which lockdown should be enforced.  Let us say, that the available amount of ventilator is $V_{f}$  and the requirement of ventilators can be modelled using the function $V(t)$. Then, the non-pharmaceutical intervention should   be imposed to ensure $V(t)  \leq V_{f}$, for all $t$. Let,  $n$ number of critical resources be considered for lockdown considerations,  and   the requirement of $i^{\text{th}}$ critical resource can be modelled  as a function of  active cases ($A$) as, 
 \begin{equation}
     R_{i} (t)= f_{i} A(t)
 \end{equation}
 Also, availability of $i^{\text{th}}$ resource is modelled as $S_{i}(t)$. For some resources like oxygen, storage and distribution capacity could also be crucial factors.  Let storage capacity and distribution capacity is considered for  $m$ and  $p$ items, respectively.   $v_{j}$ is the storage capacity requirement of $j^{\text{th}}$ item and $V_{j}(t)$ is expected total storage capacity. Similarly, $w_{j}$ is the distribution capacity requirement of $k^{\text{th}}$ item and $W_{k}(t)$ is expected  overall expected distribution capacity.  Let $T^{l}$ is the average time required to affect the rate of active cases by imposing lockdown, that is, if lockdown is declared on today; the active cases could not be controlled for the next $T^{l}$ days.  Due to this lag, the variation of active cases could be predicted up to $T_{l}$ days with a reasonable accuracy using time series analysis of active cases. It is observed that $T_{l}$ could be considered as 14 days for preliminary analysis. Let $t_{c}$ be the current time, and  $\delta$ be the time after which lockdown will be imposed. For simplicity, it is assumed that the economic value of a normal day is $\alpha$ and during lockdown it becomes zero. Then the lockdown criteria are formulated using the following cost function ($J$).

 \begin{equation}
    \text{Maximize} \quad J= \alpha \delta
 \end{equation}
 
\noindent subject  to the following conditions: \\
\noindent For sufficient availability of each of $n $ resources,  
 \begin{equation}
     f_{i} A(t) \leq S_{i}(t) ;  \quad  t \epsilon  [t_{c}, t_{c}+ T^{l}+\delta]   \label{co1}
 \end{equation}
 
\noindent  For sufficient storage of each of  $m$ resources, 
  \begin{equation}
     f_{j} A(t)v_{j} \leq V_{j}(t) ; \quad  t \epsilon  [t_{c}, t_{c}+T^{l}+\delta]  \label{storage_1}
 \end{equation}
 
\noindent   For sufficient distribution capacity of each of $n$ resources, 
  \begin{equation}
     f_{k} A(t)w_{k} \leq W_{k}(t) ; \quad  t \epsilon  [t_{c}, t_{c}+T^{l}+\delta]  \label{distribution_1}
 \end{equation}
 
\noindent  The daily growth rate $(G(t))$ should not exceed $x$ percent. 
  \begin{equation}
      G(t) \leq x  \quad  t \epsilon  [t_{c}, t_{c}+\delta]   \label{growthrate}
  \end{equation}
  
\noindent Test positive ratio $(TP(t))$ should not exceed $y$ percent. 

\begin{equation}
    TP(t) \leq y  \quad  t \epsilon  [t_{c}, t_{c}+\delta]  \label{TPR}
\end{equation}
  
  Authorities will want to provide some time to citizens to prepare for lockdown as an immediate declaration of lockdown might create chaos in social life.  In the cost function, $\delta$ represents that time, where authorities can decide for the lockdown or can warn citizens about the probable lockdown at $\delta$ time prior to actual implementation. Here, $n$ resources include the medical infrastructure like hospital beds and medical items like medicines and oxygen. The value of $x$ and $y$  can be chosen by the authority based on the situation. It is preferable to keep the maximum value of $x$ as 5 \% and $y$ as 10 \%.

 \subsection{Prediction of active cases}
 Prediction of active cases is performed using time series analysis based on the reported active cases for the last two months \citep{jana2020adaptive}. It is observed that the total number of active cases can be presented as a fourth-order polynomial function of time. Therefore, the total number of active cases can be expressed as
 \begin{equation}
     A(t)= a_{A} t^{4}+  b_{A}t^{3} + c_{A}t^{2} + d_{A}t +e_{A}  \label{act_polynomial}
 \end{equation}
where, the  co-efficients  are obtained  by minimizing the error between the observed active cases and predicted value of active cases using (\ref{act_polynomial}). Let $ A(t)= X a = \sum_{j=1}^{j=5} X_{jt}a_{j}$, where $a_{j}$ are the co-efficients. The co-efficents are obtained after minimizing the following functions,
\begin{equation}
\min_{a} \quad J^{a}=   \sum_{t=t_{0}}^{t=t_{f}}{ w_{t}(y_{t}-  \sum_{j=1}^{j=5} X_{jt}a_{j})^2} \label{wtpseudoinv}
\end{equation}
 where, $y_{t}$ is the observed value of active cases, and $w_{t}$ is the weights. Let  W  be  the diagonal matrix of weights. Then the optimal value of  $a$,  $a^{\text{opt}}$ is  derived as, 
 
 \begin{equation}
  a^{\text{opt}}= (X^{T}WX)^{-1}X^{T}Wy_{t}
 \end{equation}
 
 Therefore, the total actives  cases are predicted using the following equation, 
 
 \begin{equation}
     A(t)=X a^{\text{opt}}
 \end{equation}

 \section {Case study} 
 
In this section, a case study for Delhi is considered to validate the proposed formulation. In the case of Delhi, there was a shortage of oxygen even after the deployment of lockdown. For simplicity, we will consider only oxygen resources. The time-series data of confirmed, recovered, and death cases are obtained from "https://api.covid19india.org/" website.  

As per the Ministry of Home Affairs of Government of India order 40-3/2020-DM-I(A) dated 22.04.2021, the Delhi government is allocated 480 MT against the demand of 700  MT (considering demand on  $20^{\text{th}}$ April onwards).  Considering the total availability of 480 MT oxygen and demand of 700 MT oxygen as a current requirement of Delhi on $20^{\text{th}}$ April, we will analyze the optimal lockdown criteria as on  $6^{\text{th}}$ April. The seven days average TPR on 6th April for Delhi was 4.3 \%, and seven days average daily growth was 0.52 \% .  The total number of active cases of Delhi on $20^{\text{th}}$ April is 85575. We could assume reasonably that the demand for oxygen is proportional to the total number of active cases.  Therefore, the average demand for oxygen per active case is 0.00817 MT. So, for oxygen resources, the value of $f_{1}$ is  0.00817 MT. It is to be noted that the value of  $f_{1}$ can vary from region to region.  Prediction of active cases on $6^{\text{th}}$  April is obtained as, 
 \begin{equation}
     A(t)= a^{1}_{A} t^{4}+  b^{1}_{A}t^{3} + c^{1}_{A}t^{2} + d^{1}_{A}t +e^{1}_{A} 
 \end{equation}
where, the value of $a^{1}_{A}$, $b^{1}_{A}$, $c^{1}_{A}$, $d^{1}_{A}$, $e^{1}_{A}$ are   0.006 , -0.518, 16.088, -175.330, 1344.983. . So, the predicted number of total active cases after $t_{d}$ days from $6^{\text{th}}$ April is,

\begin{equation}
    A(t)= a^{1}_{A} (60+t_{d})^{4}+  b^{1}_{A}(60+t_{d})^{3} + c^{1}_{A}(60+t_{d})^{2} + d^{1}_{A}(60+t_{d}) +e^{1}_{A}  
\end{equation}

\begin{figure*}[htb!]
      \centering
      \includegraphics[width=0.8\columnwidth, keepaspectratio]{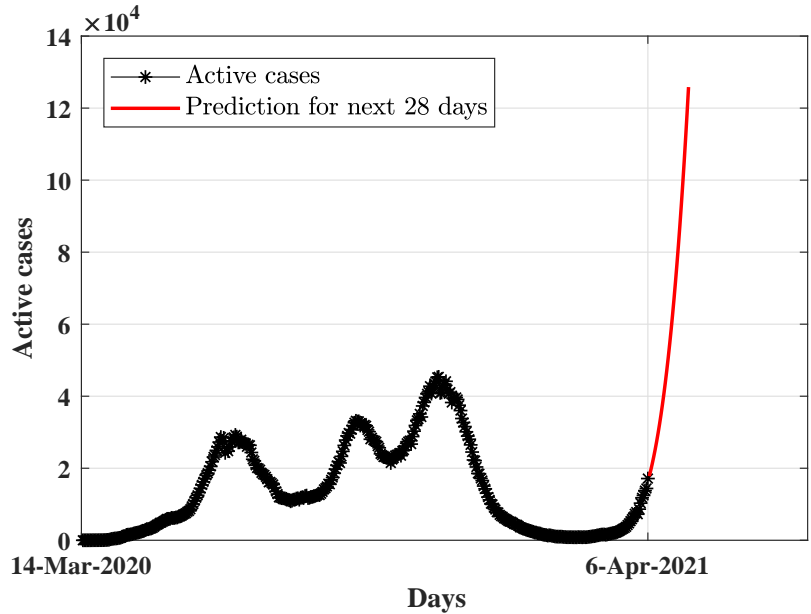}
      \caption{Prediction of Active Cases: Delhi on 6th April, 2021}
      \label{act_delhi_opt}
  \end{figure*}

\noindent The  prediction of active cases as on $6^{\text{th}}$ April is plotted in Fig. \ref{act_delhi_opt}. 

\noindent Therefore  the constraint in  (\ref{co1}) can be expressed as , 
 \begin{equation}
     0.00817 (a^{1}_{A} (t_{c}+14+\delta)^{4}+  b^{1}_{A}(t_{c}+14+\delta)^{3} + c^{1}_{A}(t_{c}+14+\delta)^{2} + d^{1}_{A}(t_{c}+14+\delta) +e^{1}_{A} ) \leq 480 \label{c1}
 \end{equation}
In this case, the previous data of the last 60 days is used for analysis; that is, the value of $t_{c}$ is 60. For simple analysis, we will assume that the storage and distribution capacity is sufficient and it is not the crucial factor.     The predicted value of  TPR and growth rate also could be modeled in a similar way. The maximum allowable value of these parameters can vary from one scenario to another.   For simplicity, we will assume that  growth rate and TPR are not the critical parameters in this case

 Therefore, the constraints in Eqns \ref{storage_1}, \ref{distribution_1}, \ref{growthrate}, and  \ref{TPR}  are not considered in the analysis. 
 The maximum value of $\delta$ which satisfies Eqn  \ref{c1} is 3.  With the value of $\delta$ equal to 3, the total requirement of oxygen could have been restricted to 480 MT.  If we have considered other constraints,  the value of $\delta$ could even be lower than 3.  Therefore, using the data  up to $6^{\text{th}}$ April, Delhi authority  should have imposed the lockdown starting from $10^{\text{th}}$ April.  In reality,  Delhi was under Curfew form $17^{\text{th}}$ April to $19^{\text{th}}$ April (Morning) and then it was under strict lockdown from evening of $19^{\text{th}}$ April.

\section{Conclusions}
In this paper,  an optimal lockdown formulation is proposed to ensure critical medical resources' availability and control the maximum value of growth rate and TPR. The proposed policy tries to optimize the duration at which lockdown could be imposed without medical crisis with available medical resources/infrastructure. The formulation is developed only using the observed value of active cases; therefore, it could be easily implementable by local authorities without considering a complex disease model.  A case study of an area with real parameters is also presented.
 \FloatBarrier

\bibliographystyle{apalike}
\bibliography{main}

\begin{thebibliography}{}

\bibitem[Acemoglu et~al., 2020]{acemoglu2020multi}
Acemoglu, D., Chernozhukov, V., Werning, I., Whinston, M.~D., et~al. (2020).
\newblock {\em A multi-risk SIR model with optimally targeted lockdown}, volume
  2020.
\newblock National Bureau of Economic Research Cambridge, MA.

\bibitem[Alvarez et~al., 2020]{alvarez2020simple}
Alvarez, F.~E., Argente, D., and Lippi, F. (2020).
\newblock A simple planning problem for {COVID}-19 lockdown.
\newblock Technical report, National Bureau of Economic Research.

\bibitem[Bandyopadhyay et~al., 2020]{bandyopadhyay2020learning}
Bandyopadhyay, S., Chatterjee, K., Das, K., Roy, J., et~al. (2020).
\newblock Learning or habit formation? optimal timing of lockdown for disease
  containment.
\newblock Technical report.

\bibitem[Barmparis and Tsironis, 2020]{barmparis2020estimating}
Barmparis, G. and Tsironis, G. (2020).
\newblock Estimating the infection horizon of {COVID}-19 in eight countries
  with a data-driven approach.
\newblock {\em Chaos, Solitons and Fractals}, page 109842.

\bibitem[Bertsimas et~al., 2021]{bertsimas2021predictions}
Bertsimas, D., Boussioux, L., Cory-Wright, R., Delarue, A., Digalakis, V.,
  Jacquillat, A., Kitane, D.~L., Lukin, G., Li, M., Mingardi, L., et~al.
  (2021).
\newblock From predictions to prescriptions: A data-driven response to
  covid-19.
\newblock {\em Health care management science}, pages 1--20.

\bibitem[Chakraborty and Ghosh, 2020]{chakraborty2020real}
Chakraborty, T. and Ghosh, I. (2020).
\newblock Real-time forecasts and risk assessment of novel coronavirus
  (covid-19) cases: A data-driven analysis.
\newblock {\em Chaos, Solitons \& Fractals}, 135:109850.

\bibitem[Dubey et~al., 2020]{dubey2020psychosocial}
Dubey, S., Biswas, P., Ghosh, R., Chatterjee, S., Dubey, M.~J., Chatterjee, S.,
  Lahiri, D., and Lavie, C.~J. (2020).
\newblock Psychosocial impact of covid-19.
\newblock {\em Diabetes \& Metabolic Syndrome: Clinical Research \& Reviews},
  14(5):779--788.

\bibitem[Federico and Ferrari, 2021]{federico2021taming}
Federico, S. and Ferrari, G. (2021).
\newblock Taming the spread of an epidemic by lockdown policies.
\newblock {\em Journal of mathematical economics}, 93:102453.

\bibitem[Ghosh et~al., 2021]{ghosh2021mathematical}
Ghosh, A., Roy, S., Mondal, H., Biswas, S., and Bose, R. (2021).
\newblock Mathematical modelling for decision making of lockdown during
  covid-19.
\newblock {\em Applied Intelligence}, pages 1--17.

\bibitem[Gonzalez-Eiras and Niepelt, 2020]{gonzalez2020optimal}
Gonzalez-Eiras, M. and Niepelt, D. (2020).
\newblock On the optimal'lockdown'during an epidemic.

\bibitem[{Gupta} and {Shankar}, 2020]{gupta2020estimating}
{Gupta}, S. and {Shankar}, R. (2020).
\newblock Estimating the number of {COVID}-19 infections in indian hot-spots
  using fatality data.
\newblock {\em arXiv preprint arXiv:2004.04025}.

\bibitem[Hikmawati et~al., 2021]{hikmawati2021multi}
Hikmawati, E., Ulfa~Maulidevi, N., and Surendro, K. (2021).
\newblock Multi-criteria recommender system model for lockdown decision of
  {COVID}-19.
\newblock In {\em 2021 10th International Conference on Software and Computer
  Applications}, pages 39--44.

\bibitem[Huang et~al., 2020]{huang2020global}
Huang, J., Zhang, L., Liu, X., Wei, Y., Liu, C., Lian, X., Huang, Z., Chou, J.,
  Liu, X., Li, X., et~al. (2020).
\newblock Global prediction system for covid-19 pandemic.
\newblock {\em Science bulletin}, 65(22):1884.

\bibitem[Jana and Ghose, 2020]{jana2020adaptive}
Jana, S. and Ghose, D. (2020).
\newblock Adaptive short term {COVID}-19 prediction for india.
\newblock {\em MedRxiv}.

\bibitem[Kumar and Choudhury, 2021]{kumar2021migrant}
Kumar, S. and Choudhury, S. (2021).
\newblock Migrant workers and human rights: A critical study on india’s
  covid-19 lockdown policy.
\newblock {\em Social Sciences \& Humanities Open}, 3(1):100130.

\bibitem[Li et~al., 2020]{li2020novel}
Li, C., Romagnani, P., and Anders, H.-J. (2020).
\newblock Novel criteria for when and how to exit a {COVID}-19 pandemic
  lockdown.
\newblock {\em Frontiers in big data}, 3.

\bibitem[Miclo et~al., 2020]{miclo2020optimal}
Miclo, L., Spiro, D., and Weibull, J.~W. (2020).
\newblock Optimal epidemic suppression under an icu constraint.

\bibitem[Mottaleb et~al., 2020]{mottaleb2020covid}
Mottaleb, K.~A., Mainuddin, M., and Sonobe, T. (2020).
\newblock Covid-19 induced economic loss and ensuring food security for
  vulnerable groups: Policy implications from bangladesh.
\newblock {\em PloS one}, 15(10):e0240709.

\bibitem[Rajesh et~al., 2020]{rajesh2020covid}
Rajesh, A., Pai, H., Roy, V., Samanta, S., and Ghosh, S. (2020).
\newblock Covid-19 prediction for india from the existing data and sir (d)
  model study.
\newblock {\em medRxiv}.

\bibitem[Rawson et~al., 2020]{rawson2020and}
Rawson, T., Brewer, T., Veltcheva, D., Huntingford, C., and Bonsall, M.~B.
  (2020).
\newblock How and when to end the {COVID}-19 lockdown: an optimization
  approach.
\newblock {\em Frontiers in Public Health}, 8:262.

\bibitem[Ren, 2020]{ren2020pandemic}
Ren, X. (2020).
\newblock Pandemic and lockdown: a territorial approach to covid-19 in china,
  italy and the united states.
\newblock {\em Eurasian Geography and Economics}, 61(4-5):423--434.

\bibitem[Singh et~al., 2020]{singh2020income}
Singh, I., Singh, J., and Baruah, A. (2020).
\newblock Income and employment changes under covid-19 lockdown: A study of
  urban punjab.
\newblock {\em Millennial Asia}, 11(3):391--412.

\bibitem[Usher et~al., 2021]{usher2021covid}
Usher, K., Bradbury~Jones, C., Bhullar, N., Durkin, D.~J., Gyamfi, N., Fatema,
  S.~R., and Jackson, D. (2021).
\newblock Covid-19 and family violence: Is this a perfect storm?
\newblock {\em International journal of mental health nursing}.

\end{thebibliography}

\end{document}